\documentclass[11pt]{article}
\usepackage{amssymb}
\usepackage{euscript}

\begin{document}

\newcommand{\be}{\begin{equation}}
\newcommand{\ee}{\end{equation}}
\newcommand{\epe}{\end{equation}}
\newcommand{\bea}{\begin{eqnarray}}
\newcommand{\eea}{\end{eqnarray}}
\newcommand{\ba}{\begin{eqnarray*}}
\newcommand{\ea}{\end{eqnarray*}}
\newcommand{\epa}{\end{eqnarray*}}
\newcommand{\ar}{\rightarrow}

\def\s{\sigma}
\def\r{\rho}
\def\D{\Delta}
\def\R{I\!\!R}
\def\l{\lambda}
\def\g{\gamma}
\def\D{\Delta}
\def\cD{{\cal D}}
\def\cH{{\cal H}}
\def\dA{A^{\dag}}
\def\d{\delta}
\def\T{\tilde{t}}
\def\X{\tilde{X}}
\def\ts{\tilde{\sigma}}
\def\k{\kappa}
\def\t{\tau}
\def\G{\Gamma}
\def\f{\phi}
\def\p{\psi}
\def\z{\zeta}
\def\ep{\epsilon}
\def\hx{\widehat{\xi}}
\def\A{\tilde{a}}
\def\B{\tilde{b}}
\def\a{\alpha}
\def\b{\beta}
\def\O{\Omega}
\def\H{\cal H}
\def\tH{\tilde{H}}
\def\M{\cal M}
\def\g{\hat g}
\newcommand{\dslash}{\partial\!\!\!/}
\newcommand{\aslash}{a\!\!\!/}
\newcommand{\eslash}{e\!\!\!/}
\newcommand{\bslash}{b\!\!\!/}
\newcommand{\vslash}{v\!\!\!/}
\newcommand{\rslash}{r\!\!\!/}
\newcommand{\cslash}{c\!\!\!/}
\newcommand{\fslash}{f\!\!\!/}
\newcommand{\Dslash}{D\!\!\!\!/}
\newcommand{\Aslash}{{\cal A}\!\!\!\!/}


\vspace{1cm}
\begin{center}

{\LARGE String Entanglement and D-branes as Pure States}

\vspace*{1cm} {\large Marcelo Botta Cantcheff $^{\ddag}$
\footnote{e-mail: botta@fisica.unlp.edu.ar, botta@cbpf.br}}

\vspace{3mm}

$^{\ddag}$ {\it IFLP - CONICET} \, \textit{and} \,{\it Departamento de F\'\i sica, UNLP, \\\it cc67, CP1900 La Plata, Argentina}

\end{center}

\begin{abstract}
\noindent

We study the entanglement of closed strings degrees of freedom in
order to investigate the microscopic structure and statistics of
objects as D-branes. By considering the macroscopic pure state (MPS)
limit, whenever the entanglement entropy goes to zero (in such a way that
the macroscopic properties of the state are preserved), we show that
boundary states may be recovered in this limit and, furthermore, the
description through closed string (perturbative) degrees of freedom
collapses. We also show how the thermal properties of branes and
closed strings could be described by this model, and it requires
that dissipative effects be taken into account. Extensions of the MPS
analysis to more general systems at finite temperature are finally
emphasized.

\end{abstract}


\section{Introduction}

Since the first microscopical description of the black hole
entropy \cite{bh}, one of the most interesting problems concerning
D-branes is the development of a model where their thermodynamical
properties and microscopical structure may be clarified. One may
study features of D-branes at finite temperature by using
dualities \cite{vazquez}, however a detailed statistical model
should start from the understanding of the true microscopical
degrees of freedom that describe the brane.

Closed strings are believed to describe the perturbative degrees of freedom of a
 quantum theory that includes gravity, and D-branes are macroscopical objects which
  absorb or emit closed strings \cite{boundarystates}; on the other hand,
   these ones are identified with classical gravitational solitons that can work as
    backgrounds in a perturbative scheme \cite{RSZ}.
At present, there is no known theory in which the D-branes are described by vacuumlike
 states or by states created from a vacuum. Apart from this, any satisfactory description
  in this direction should include thermal effects, since generic nontrivial background
   spacetimes
carry thermodynamic properties, such as temperature and entropy. A recent approach
 realizes these properties \cite{brane-tfd}, but it is based on an open string description
  whose precise relation with the graviton excitations is unknown. Other models to describe
   D-branes at finite temperature, like boundary states of thermal closed strings came up
    in Refs. \cite{IVV}, these states have,
however, nothing to do with vacuum configurations in any sense.

This work is devoted to a two fold purpose: to shed light on all these subjects by
studying the general entanglement of the right/left sectors of closed strings;
and, by doing that, to show how the boundary state that describes the D-brane may be
 obtained from statistical analysis of the pure macroscopic states without
  using any link to string boundary conditions. We actually show that boundary
   states may be recovered as macroscopic pure states in a vanishing entropy limit,
    and we see furthermore that novel (nonperturbative) degrees of freedom are
     needed to describe the
dynamics of these objects.

There are examples in which the entanglement of the degrees of
freedom of single closed strings is dynamically generated in backgrounds with
 gravitational fields \cite{ent-grav1, hor, ent-grav2}. This is the main motivation
  to study more deeply this behavior and to highlight the appearance of
structures like the D-branes.

The paper is organized as follows. In Section 2, we briefly review
the boundary state formalism. In Section 3, we study the general
entanglement of the internal closed string modes and, by using the
canonical symmetries, define canonical variables adapted to the
$p+1$ dimension of the brane. The main results are found in
Section 4, where (coherent) macroscopical states, localized in the
brane surface are built up as fundamental states in a closed
string Hilbert space. Furthermore, these states are proposed to
describe the statistical properties of the brane, the entropy
operator is canonically defined, and boundary states are recovered
as pure states. The finite temperature behavior of branes (and
closed strings being created in them) is also discussed. In
Section 5, the pure state limit is analyzed and its generalization
to other thermodynamic systems is pointed out. Finally, in Section
6, we summarize the main conclusions of this approach.

\section{Closed strings and Boundary States}

Let us consider a closed bosonic string in the Minkowski space-time
 $(\R^{26}, \eta_{\mu \nu
})$. The general solution with periodical boundary conditions
reads:\be X^\mu = x^\mu_0 - i\a' p^\mu t +
\sqrt{\frac{\a^\prime}{2}}\sum_{n > 0}\frac{1}{\sqrt{n}}
\bigg[\,(\a_n^\mu e^{-i\,n\,(t-\s \,)} + \a_n^{\dagger \:
\mu}e^{i\,n\,(t-\s\,)}\,) + \,({\b}_{n}^{\mu} e^{-i\,n(t+\s)}
+{\b}_{n}^{\dagger \: \mu} e^{i\,n(t + \s)}\,)\bigg], \ee
 After its light-cone gauge
quantization (LCG), Fourier coefficients $\alpha ^{I}_{n}$ and $\beta ^{I}_{n}$
 ($I,J,K=1,...24$ denote the coordinates transverse to $X^{\pm}= X^{0} \pm X^{25}$),
  for the left- and right-moving modes respectively, can be
redefined in order to obtain the physical creation and annihilation operators for
each mode $n$ in the different sectors. Namely,
\begin{eqnarray}
A_{n}^{I} &=&\frac{1}{\sqrt{n}}\alpha _{n}^{I},
\qquad A_{n}^{I\dagger}
=\frac{1}{\sqrt{n}}\alpha_{-n}^{i}, ~~~\forall~ n~>~0,\\
B_{n}^{I} &=&\frac{1}{\sqrt{n}}\beta _{n}^{I},
\qquad B_{n}^{I\dagger}=\frac{1}{\sqrt{n}}\beta _{-n}^{I},~~~\forall~ n~>~0.
\end{eqnarray}
 These redefined operators satisfy the oscillator-like
canonical commutation relations (CCR):
\begin{eqnarray}
\left[ A_{n}^{I },A_{m}^{J \dagger }\right]  &=&\left[ B
_{n}^{\mu I}, B_{m}^{\nu J \dagger }\right] =\delta _{nm}\eta^{IJ},  \nonumber \\
\left[ A_{n}^{I },B_{m}^{J}\right]  &=&\left[ A_{n}^{I },
B_{m}^{J \dagger }\right]  =...=0.
\label{opalg}
\end{eqnarray}

 The fundamental state of the closed
bosonic string is defined by
\begin{equation}
A_{n}^{I}\left| 0\right\rangle = B_{n}^{I}\left| 0\right\rangle
=0,
\end{equation}
where $\left| 0\right\rangle =\left| p\right\rangle\left|
0\right\rangle _{A }\left| 0\right\rangle _{B } $, as usual. The
momentum of the zero mode is often defined $p \equiv 0$, however
it is actually arbitrary by virtue of the relativistic invariance of
the string vacuum. The Dp-brane states are given by the following
operatorial equations on states of the one-closed string Hilbert
space, built from the usual boundary state conditions,
\be\label{B1a} (A^{a}_n + B^{\dagger\:a}_n
)\;\left|B_p\right\rangle = 0 , \;\;\; a=0,1,..., p\ee
\be\label{B1i} (A^{i}_n - B^{\dagger\:i}_n )\;\left|B_{p}
\right\rangle = 0 , \;\;\; i=p+1,...25\ee and for the zero mode we
get \be \label{B3}  p^a \;\;\left|B_{p} \right\rangle = 0 ,\ee \be
\label{B2} (x_{0}^i -  x^i )  \;\left|B_{p} \right\rangle = 0 ,
\ee
 where $x^i$ are the coordinates of the Dp-brane hyperplane. The solution reads as
\be
\label{boundary} \left|B_{p}
\right\rangle = C_{p}
\delta(x^i_{0} - x^i) \prod_{I, n>0} \,\,
 e^{-\left(S_p\right)_{IJ} A^{\dagger\:I}_{n} B^{\dagger\:J}_{n}} \left|0\,\!\right\rangle \, ,
 \ee
where we have defined $\left(S_p\right)^I_J \equiv (\d^a_b, -\d^i_j )$,
 $C_{p}$ is a normalizing factor related to the
brane tension by $T_p = 2C_{p}$.

The Hamiltonian operator writes
\be
H ~=~\sum_{n >0}^{\infty }
n\left( A_{n}^{\dagger }\cdot A_{n}+ B_{n}^{\dagger }\cdot B_{n} + tr\,\eta^{IJ}  \right) ~\equiv~\sum_{n >0}^{\infty }
\,n \, H_n ,
\label{exthamilt}
\ee
where the dot represents an Euclidean scalar product in the
transverse space, $a \cdot b\equiv \eta_{IJ}a^I b^J $. The negative quantity
\be
V\equiv \sum_{n >0}^{\infty } \,n \,tr\,\eta^{IJ}
\ee
is the energy of the tachyon in the rest-frame.

To construct the physical Fock space it is necessary to fix the
residual gauge symmetry generated by the world sheet canonical momentum $\Pi$.
This imposes the level matching condition (LMC) on a physical
state $\left |\Psi\right\rangle$:
\begin{equation}
\Pi\left |\Psi\right\rangle= \sum_{n=1}^{\infty }\left( nA_{n}^{\dagger }\cdot
A_{n}- nB_{n}^{\dagger }\cdot B_{n}\right)\left|\Psi\right\rangle= \sum_{n=1}^{\infty }\,n\left(N^A_n - N^B_n \right)\left|\Psi\right\rangle=0,
\label{lmc}
\end{equation}
  where we have defined the number operators $N^A_n \equiv A_{n}^{\dagger }\cdot
A_{n}$, $N^B_n \equiv B_{n}^{\dagger }\cdot
B_{n}$.

   \section{String entanglement and generalized vacua}

In what follows we study the internal quantum entanglement of closed strings, which may be reduced to entanglement between the right/left
 sectors. According to previous evidence \cite{ent-grav1, hor, ent-grav2}, this behavior is generally induced
  in nontrivial backgrounds. Here, we will assume this generically, in order to study the consistency with expected statistic properties
  in gravitational backgrounds, so as other implications in the theory of D-branes.

  The entanglement between two independent parts
    of a system is often described by a Bogoliubov transformation acting on the
     Hilbert space, which must be a tensor product of the two states space of each
      subsystem \cite{vitiello}. In this case we have
\be
e^{iG(\theta)}:\, {\cal
H}_A \otimes {\cal H}_B \, \,  \,\mapsto \,\, {\cal
H}_A \otimes {\cal H}_B\,\,\,\,,
\ee
where $G(\theta)$ is the generator of the transformation, called
{\em the Bogoliubov operator}, and $\theta$ represents the set of parameters
 of the transformation, which shall depend on external conditions that induce
  the entanglement \cite{vitiello}. The pure closed string vacuum
   $\left| 0\right\rangle =\left| p \right\rangle\,\left| 0\right\rangle
_{A}\left| 0\right\rangle _{B} $, is transformed into a coherent mixed state
\be\label{entangled-closed-vacuum}\left|0(\theta)\right\rangle\
 = e^{iG(\theta)}\left| 0\right\rangle
 \, ,\ee
which is annihilated by the transformed operators $A_n(\theta)$
and $ B_n(\theta)$. The zero mode $\left| p \right\rangle$ is
independent of the entangled sector.

In this framework one canonically quantizes the
 fields as operators and the statistical average of an operator
 $Q$ is defined as its expectation value
in the entangled vacuum state (\ref{entangled-closed-vacuum}). So this state
 encodes the quantum and statistical information of the system on behalf of a
  density operator.

The most basic feature that characterizes a Bogoliubov
transformation is that it preserves the canonical commutation
relations. This reflects the fact that the original nature of the
degrees of freedom is preserved, even if the effective dynamics is
different. Thus, it shall be assumed that this map preserves both
the CCR and the level matching condition in order to ensure that
the transformed degrees of freedom are also closed strings; thus
in particular, the generator $G$ commutes with the operator $\Pi$.
We will also assume that this map is unitary in order to preserve
the amplitudes and probability measures.

 The general form of the Bogoliubov transformation that fix
the form of the generator is given by the following relation \cite{ume1,chu-ume}
\begin{equation}
\left(
\begin{array}{c}
A^{\prime } \\
B^{\dagger \prime }
\end{array}
\right) =e^{-iG}\left(
\begin{array}{c}
A \\
B^{\dagger }
\end{array}
\right) e^{iG}={\cal B}\left(
\begin{array}{c}
A \\
B^{\dagger }
\end{array}
\right) ,\quad \left(
\begin{array}{cc}
A^{\dagger ^{\prime }} & -B^{\prime }
\end{array}
\right) =\left(
\begin{array}{cc}
A^{\dagger } & -B
\end{array}
\right) {\cal B}^{-1},
\label{Bogoliubov}
\end{equation}
where $G$ is Hermitian, and then $\cal B $ is a $2 \times 2$ complex matrix
\begin{equation}
{\cal B}=\left(
\begin{array}{cc}
u & v \\
v^{*} & u^{*}
\end{array}
\right) ,\ee
such that
\be
\qquad \left| u\right| ^{2}-\left| v\right| ^{2}=1,
\label{Bmatrix}.
\end{equation}
The relation (\ref{Bmatrix}) encodes the preservation of the CCR and it will be
 important for our interpretation of the results.
The operators
that satisfy the relations (\ref{Bogoliubov}) and (\ref{Bmatrix}) have the
following form \cite{chu-ume}
\begin{eqnarray}
G_{1_{n}}^{} &=&\theta _{1_{n}}\left( A_{n}\cdot B_{n}+ B_{n}^{\dagger }\cdot A_{n}^{\dagger }\right) ,  \nonumber \\
G_{2_{n}}^{} &=&i\theta _{2_{n}}\left( A_{n}\cdot B_{n}-
B_{n}^{\dagger }\cdot A_{n}^{\dagger }\right) , \nonumber \\
G_{3_{n}}^{} &=&\theta _{3_{n}}\left( A_{n}^{\dagger }\cdot A_{n}+
B_{n}^{\dagger }\cdot B_{n}+\delta_{nn}tr\eta ^{\mu \nu} \right)= \, \theta _{3_{n}}\, H_n,
\label{generators}
\end{eqnarray}
where $H_n$ in the last line, generates the time evolution of the
modes labeled by $n$. The $\theta$'s are the real parameters
which, for convenience, have been included in the operators. It is
easy to verify that the generators (\ref{generators}) satisfy the
$SU\left( 1,1 \right)$ algebra \be \left[
G_{i_{n}},G_{j_{n}}\right] =-i\Theta _{ijk}G_{k_{n}}\,\,\, ,
\label{su11} \ee where we have defined
\begin{equation}
\Theta _{ijk}\equiv 2\frac{\theta _{i_{n}}\theta _{j_{n}}}{\theta _{k_{n}}}.
\label{thetas}
\end{equation}
As we can see from  (\ref{generators}), the most general
entanglement generator $G=\sum_{n}\left( G\right)_{n}$ takes the
following form \be G_{n} =\lambda _{1_{n}}B_{n}^{\dagger }\cdot
A_{n}^{\dagger }-\lambda _{2_{n}}A_{n}\cdot B_{n}+\lambda
_{3_{n}}\left( A_{n}^{\dagger }\cdot A_{n}+B_{n}^{\dagger }\cdot
B_{n}+\delta_{nn}tr\eta ^{\mu \nu} \right) \label{rlgen} \ee and
the coefficients represent complex linear combinations of
$\theta$'s
\begin{equation}
\lambda _{1_{n}}=\theta _{1_{n}}-i\theta _{2_{n}},\quad \lambda
_{2_{n}}=-\lambda _{1_{n}}^{*},\quad \lambda _{3_{n}}=\theta _{3_{n}}.
\label{lambdas}
\end{equation}
These are called generalized Bogoliubov transformations or
simply G-transformations \cite{ume1,chu-ume} which form a $su(1,1)$ algebra.
 At this point we wish to point out that other alternatives to unitary transformations
  may be considered, although in general they lead to the same expressions for the generators
   but a different relations between these parameters \cite{ume1,chu-ume}.

By applying the
disentanglement theorem for $su( 1,1 )$ \cite{disentanglement}, one can
write the most general closed string
vacuum (\ref{entangled-closed-vacuum}) under the following form
\bea
 \left| 0\left( \theta \right) \!\right\rangle
=\prod_{n}e^{\Omega_{1_{n}}\left( B_{n}^{\dagger }\cdot
A_{n}^{\dagger }\right) }e^{\log \left( \Omega_{3_{n}}\right) \left(
A_{n}^{\dagger }\cdot A_{n}+ B_{n}^{\dagger }\cdot B
_{n}+\delta_{nn}tr\eta ^{\mu \nu}\right) }e^{\Omega_{2_{n}}
\left( A_{n}\cdot B_{n}\right)}
 \left| 0 \!\right\rangle  \;\;, \label{vacalpha}
\eea
where the coefficients of various generators are given by the relations
\be
\Omega_{1_{n}}=\frac{-\lambda _{1_{n}}\sinh \left( i\Lambda _{n}\right) }{%
\Lambda _{n}\cosh \left( i\Lambda _{n}\right) +\lambda _{3_{n}}\sinh \left(
i\Lambda _{n}\right) },\quad \Omega_{2_{n}}=\frac{\lambda _{2_{n}}\sinh
\left( i\Lambda _{n}\right) }{\Lambda _{n}\cosh \left( i\Lambda _{n}\right)
+\lambda _{3_{n}}\sinh \left( i\Lambda _{n}\right) },
\label{lambda12}
\ee
\begin{equation}
\Omega_{3_{n}}=\frac{\Lambda _{n}}{\Lambda _{n}\cosh \left( i\Lambda
_{n}\right) +\lambda _{3_{n}}\sinh \left( i\Lambda _{n}\right) },
\label{lambda3}
\end{equation}
and
\begin{equation}
\Lambda _{n}^{2}\equiv \left( \lambda _{3_{n}}^{2}+\lambda _{1_{n}}\lambda
_{2_{n}}\right) .
\label{biglambda}
\end{equation}
Since the pure vacuum is annihilated by $A_{n}^{\mu }$
and $B_{n}^{\mu }$,
the only contribution to the mixed vacuum is given by
\begin{equation}
 \left| 0(\theta )\right\rangle
=\prod_{n}(\Omega_{3_{n}})^{2\delta_{nn}tr\eta ^{\mu \nu}}
e^{\Omega_{1_{n}}\left( B_{n}^{\dagger
}\cdot A_{n}^{\dagger }\right) }  \left| 0\right\rangle\,\,\,\,.
\label{thermvacfin}
\end{equation}
The string operators are mapped to entangled ones by the
corresponding Bogoliubov generators \be A_{n}^{\mu }\left( \theta
\right)  = e^{-iG_{n}}A_{n}^{\mu }e^{iG_{n}},\qquad B_{n}^{\mu
}\left( \theta \right) =e^{-iG_{n}}B_{n}^{\mu }e^{iG_{n}}.
\label{thermop} \ee Similar relations hold for the creation
operators. The entangled operators satisfy the same canonical
commutation relations as the
 pure operators at $\theta=0$ by construction.
Alternatively, one can organize the operators in doublets \cite{ume1,chu-ume} and
 represent the Bogoliubov transformation
as
\begin{equation}
\left(
\begin{array}{c}
A_{n}^{\mu }\left( \theta \right)  \\
B_{n}^{\mu \dagger }\left( \theta \right)
\end{array}
\right) ={\cal B}_{n}\left(
\begin{array}{c}
A_{n}^{\mu } \\
B_{n}^{\mu \dagger }
\end{array}
\right) ,
\label{doublet}
\end{equation}
where the explicit form of the ${\cal B}_n$ matrices is given by
\begin{equation}
{\cal B}_{n}=\cosh \left( i\Lambda _{n}\right) {\Bbb I} +\frac{\sinh \left(
i\Lambda _{n}\right) }{\left( i\Lambda _{n}\right) }\left(
\begin{array}{cc}
i\lambda _{3_{n}} & i\lambda _{1_{n}} \\
i\lambda _{2_{n}} & -i\lambda _{3_{n}},
\end{array}
\right)
\label{explB}
\end{equation}
where $\Bbb I$ is the $2 \times 2$ identity matrix.

\vspace{0.5cm}

{\large \bf Symmetries and canonical $p$-coordinates}

\vspace{0.5cm}

This general $su(1,1)$ algebra contains the subalgebra of the canonical transformations,
 whose generators are those that
commute with the total Hamiltonian (\ref{exthamilt}):
\be\label{gen-can}
G_{3_{n}}=\theta _{3_{n}}\, H_n  \,.
\ee
These are the symmetries that may be thought of as the gauge freedoms in the present
 construction. Thus, we may use them to fix canonical coordinates encoding the information
  on the brane dimension, that consist in the original fields multiplied by relative
   real phase factors. In particular we may define this new set of annihilation operators as:
\be\label{p-coord}
A_n \,\,\to \,\,\,a_n \,\,\,\, \, ,\,\,\,\,\,\,\,B_n \,\,\to \,\,\,-\left(S_p\right)^I_J B^J_n \equiv {\tilde a}^I_n\,\,\,\,\,\,,
\ee
and the corresponding creation operators are defined by their adjoints. One may easily
 verify that in fact these transformations may be generated by operators (\ref{gen-can}).
  This fixing will clarify the interpretation of our results on $p$ branes.

The operators above annihilate the corresponding $p$-vacuum state,
\be\label{a-theta-vac} {a}^{I}_{n}\left.\left|0_p\right\rangle \!
\right\rangle= \A_{n}^I \left.\left|0_p\right\rangle \!
\right\rangle = 0
  , \ee for $n>0$ and $\left.\left|0_p\right\rangle \!
\right\rangle =|0_p\rangle \otimes|\widetilde{0}_p \rangle$ as usual, which is equivalent
 the direct product between the $A$ and $B$ vacua (up to a phase factor).
Once more, the entangled fundamental state is obtained
from this through a Bogoliubov transformation, $e^{-i{G}}$, which
mixes the two independent right/left Hilbert spaces.

The normal modes redefined above,
satisfy the canonical algebra:
\begin{eqnarray}
\left[a_{n}^I,a_{m}^{\dagger \:J}\right]
&=&\left[\tilde{a}^{I}_{n},\tilde{a}_{m}^{\dagger\:J}\right]
=  \delta_{n,m}\delta^{I,J},\label{alg}
\nonumber
\\
\left[a_{n}^{\dagger\:I},\tilde{a}_{m}^{J}\right]
&=&\left[a_{n}^{\dagger\:I},\tilde{a}_{m}^{\dagger\:J}\right]
=\left[a_{n}^{I},\tilde{a}_{m}^{J}\right]=
\left[a_{n}^{I},\tilde{a}_{m}^{\dagger\:J}\right]=0.
\end{eqnarray}

Let us notice finally that a transformation of this type does not
generate any contribution to the entanglement entropy. In fact it
does not mix between right- and left-modes, by virtue of the
separated structure of the generators
$G_{3_{n}}=G_{3_{n},A}+G_{3_{n,B}}$ (the corresponding matrices
${\cal B}_{(3)\, n}$ are indeed diagonal).

\section{Coherent States: Entropy and Brane Thermodynamics}

In order to keep the fixing above we shall consider a subgroup which led the
 $p$-coordinates choice invariant.
In most of the systems, the thermal effects are described by
standard one-parameter
 Bogoliubov transformations,
generated by $g _{2_{n}}$ \cite{ume2,ume4}. In what follows, we focus on these
 type of transformations
since one of our purposes is to describe those effects.

\vspace{0.5cm}

{\large\bf Vacuum/boundary states}

\vspace{0.5cm}

In terms of the new canonical variables, standard unitary Bogoliubov transformations
 are generated by: \be\label{g2} G(\theta)= -i
\,\delta_{I J}\,\sum_{n}\theta_{n}\,(a_{n}^{I} \tilde{
a}_{n}^{J}\, - \,\tilde{a}_{n}^{\dagger\:I} a_{n}^{\dagger\:J} )\, ,
\ee
for finite volume systems. The creation and annihilation are transformed according
to
\bea a_{n}^{I}(\theta_{n}) &=& e^{-iG}a_{n}^{I}e^{iG}
=\cosh(\theta_{n})a_{n}^{I} - \sinh(\theta_{n}){\widetilde
a}_{n}^{\dagger \: I}
\\
\widetilde{a}_{n}^{ I}(\theta_{n}) &=& e^{-iG}{\widetilde
a}_{n}^{ I}e^{iG}
= \cosh(\theta_{n}) a_{n}^{I} -  \sinh(\theta_{n}) {\widetilde
a}_{n}^{\dagger \: I}\,\,.
\eea
These operators annihilate the states:
\begin{eqnarray}
a_{n}^{I}(\theta_{n})\left |0(\theta)\right\rangle &=& \widetilde
{a}_{n}^{I}(\theta_{n})\left |0(\theta)\right\rangle = 0\,;
\end{eqnarray}
that, by virtue of this, must be referred to as (entangled) vacuum states.
By using the
Bogoliubov transformation, these relations give rise to the vacuum
state conditions:
\begin{eqnarray}
\left[a_{n}^{I}-
\tanh(\theta_n)\widetilde{a}^{\dagger \: I}_n\right]
\left|0(\theta)\right\rangle
&=&0,
\label{cond1}
\\
\left[\widetilde{a}_{n}^{I}-
\tanh(\theta_n){a}^{\dagger\: I}_{n}\right]
\left|0(\theta)\right\rangle
&=&0,\label{cond2}
\end{eqnarray}
which are also generalizations of the boundary conditions (\ref{B3}), (\ref{B2}) \footnote{These ones may naturally be associated to ${(\theta)}$-deformed closed string solutions $X^I_{(\theta)}(t,\sigma)$.}. So the present framework handles a consistent two-fold interpretation for these states: as ground states, and also as (deformed) boundary states.

The solution of (\ref{cond1}) and (\ref{cond2}) is \be\left. \left| 0_p(\theta )\right\rangle \!\right\rangle  = e^{-i{G}}
\left.\left|0_p\right\rangle \! \right\rangle = \,\delta_p\, \prod_{n=1}\left[\left( \frac{1}{\cosh(\theta_{n})}\right)^{D-2}
e^{\tanh(\theta_{n})\,\delta_{I J}\,a_{n}^{\dagger\:I} {\tilde
a}_{n}^{\dagger\:J}} \right]
\left.\left|0_p\right\rangle\!\right\rangle \label{tva} ; \ee
and in terms of the $A$/$B$ modes, it expresses in the suggestive form:
\begin{equation}\label{entangled-boundary}
\left |0_p(\theta)\right\rangle
=\,\delta_p\,\prod_{n>0} \,\left( \frac{1}{\cosh(\theta_{n})}\right)^{D-2}\,
e^{-  \tanh(\theta_{n})\,  \left(S_p\right)_{IJ} A^{\dagger\:I}_{n} B^{\dagger\:J}_{n}} \left| 0\right\rangle .
\end{equation}
By considering an ensemble of closed strings in their fundamental state ($p^i=0\,,\, \,p^a=0 $) one may construct coherent states localized (so as a wave packet) in a $p$-dimensional surface $x^i=const$. These vacua are all solution of (\ref{cond1}),(\ref{cond2}), and are determined by zero mode conditions (\ref{B3})(\ref{B2}). This is expressed by the prefactor $\delta_p\equiv\delta(x^i_0 - x^i)$ in the expressions above ((\ref{tva}), (\ref{entangled-boundary})). Notice then that these generalized $p$-vacua are coherent states with macroscopic properties localized in the brane hypersurface.

\vspace{0.5cm}

Then, the Fock space of entangled  string states is constructed by
applying the mixed creation operators
$\,a_{n}^{\dagger\:I}(\theta) \, ,\, \,{\tilde a}_{n}^{\dagger\:I}
(\theta)\,$, to the vacuum $(\ref{tva})$
 that may properly be identified with a brane state (because of the condensate
  structure in itself) absorbing and emitting entangled closed strings.
 In fact, if the modes corresponding to the $(\theta)$-string leaving the
  $D_{(\theta)}$-brane are created from this ground state $(\ref{tva})$,
 then in absence of such excitations, the $D_{(\theta)}$-brane on its own must be
  associated to the fundamental state.

Remarkably enough, (\ref{cond1}) and (\ref{cond2}) are a generalization of
 the boundary conditions (\ref{B1a}) and (\ref{B1i})
  (to be recovered in the proper limit), and in this sense, this state
   is a natural candidate to describe an statistical ensemble associated to the
    $D_p$-brane, which handles its thermal properties. The following part is devoted to
the study of this issue, and it is addressed to recover $D_p$-branes as
 ground states.

\vspace{0.5cm}

{\large\bf Statistic Analysis}

\vspace{0.5cm}

The entanglement entropy operator is defined such that its average value be proportional to
the thermodynamical
entropy of any free bosonic field divided by the Boltzmann's
constant
\cite{ume2}. For a bosonic field, it can be computed
as the
expectation value of the entropy operator in the state that describes the system
\begin{equation}\label{entrboson}
S\equiv\frac{1}{k_{B}}\left\langle
 K \right\rangle
= -\left\{ \sum_{k}\left[{\cal N}_{k}\log  {\cal N}_{k} -\left( 1+
{\cal N}_{k}\right) \log \left( 1+{\cal N}_{k}\right)\right]
\right\}\,\,\,,
\end{equation}
where ${\cal N}_k$ is the number of particles in the state $k$. This general expression
 is nothing but the expectation value of the von Neumann entropy operator
  $ K \equiv -\, {\cal N}\, \log  {\cal N} $, where the second term has taken into account
   the counting of antiparticles states \cite{vitiello}.
Consequently, one defines the
entropy operator for the bosonic string through the number operators $N_n^{A/B}$ for
 general $SU(1,1)$ transformations (\ref{rlgen}) according to this formula.
The LMC on physical states (Eq. (\ref{lmc}))  implies that $ K_n^{A} = K_n^{B}$, so the
relevant quantity for our analysis is the entropy associated with one of the two (right/left)
 sectors. In the particular case of standard transformations generated by (\ref{g2}), the
  number operator of left-handed modes is proportional to $\sinh^2
\theta_n$; therefore, in terms of canonical $p$-coordinates, the associated entropy reduces to
\begin{eqnarray}
 K_{} &=& -\,\sum_{n=1} \bigg\{ a_{n}^{\dagger \: I} a^J_{n } \,\delta_{I J}\,\ln \left(
\sinh^{2}\left(\theta _{n}\right)\right) - a^I_{n} a_{n}^{\dagger \: J
} \,\delta_{I J}\, \ln \left( \cosh^{2}\left(\theta _{n}\right)\right)\bigg\} .
\label{k}
\end{eqnarray}

A straightforward algebra leads to the following expression for the
vacuum state
      \be\label{boundarytermico-entropy}
  \left.\left|0_p(\theta)\right\rangle\!\right\rangle=\, \delta_p\,\,e^{-K(\theta)/2}\,\,\prod_{I, n>0}
\,\, e^{ a^{\dagger\:I}_{n} \A^{\dagger\:I}_{n}}
\left.\left|0_p\right\rangle \!\right\rangle=\, \delta_p\,
\,e^{-\tilde{K}(\theta)/2}\,\,\prod_{I, n>0} \,\, e^{
a^{\dagger\:I}_{n} \A^{\dagger\:I}_{n}}
\left.\left|0_p\right\rangle\!\right\rangle ,\ee in terms of the
entropy operator \footnote{A demonstration of this expression may
be found in Ref.\cite{vitiello}, where the entanglement of quantum
fields in presence of event horizons is studied.}. Then, according
to (\ref{boundary}), it may be written as:
       \be\label{boundarytermico-entropy2}
  \left.\left|0_p(\theta)\right\rangle\!\right\rangle=\,\frac{1}{C_{p}} \,e^{-K(\theta)/2}\, \,
\left|B_p\right\rangle =\, \frac{1}{C_{p}}\,e^{-\tilde{K}(\theta)/2}\, \,
\left|B_p\right\rangle\, .\ee
This expression shows how the ground state of a system of
 self-entangled strings is related to the $D_p$-brane state through the
  entropy operator, and they both coincide in the formal limit $K\to 0$. This limit
   will be analyzed in the next section, and may be seen as an alternative way to
    construct the boundary state (\ref{boundary}).

A very important remark has to be done at this point. Recalling
the generalized vacuum state conditions ((\ref{cond1}),
(\ref{cond2})), the boundary state (\ref{boundary}) may be
recovered as a vacuum state or, more precisely, as a proper
(vanishing entropy) limit point of a many-string ground state.

Therefore the generalized $p$-vacua (expressed by (\ref{entangled-boundary})) may be
 seen as statistical or thermodynamical extensions (parametrized by $\theta$) of the
  $D_p$-brane state, since this is a coherent state of entangled string modes localized
   on the D-brane surface. 
In this sense, the brane tension may be promoted to an operator that act on the pure
 brane state, defined in terms of the brane entropy:
\be\label{tension}
\hat{T}_p(\theta)\equiv \,2\,e^{-K(\theta)/2}\,
\ee
then the state that represents the \emph{statistical brane} expresses as
\be
\left.\left|0_p(\theta)\right\rangle\!\right\rangle=\, \frac12\,\delta_p\,\,\,\,\,\hat{T}_p(\theta)\,\,\prod_{I, n>0}
\,\, e^{ a^{\dagger\:I}_{n} \A^{\dagger\:I}_{n}} \,\,\,\left.\left|0_p\right\rangle\!\right\rangle\,\,.
\ee

 $D$-branes are believed to be solitons in (super)-gravity theories that work
  as the ground states for gravitons (carried by closed strings). In agreement
   with this fact, the calculation above confirm that $D$-branes configurations
    indeed appear under general conditions of entanglement of closed string fields;
     provided that these entanglement effects arise in gravitational backgrounds
      \cite{ent-grav1, hor, ent-grav2, vitiello}. Then as a by-product, if one associates
       the entropy $K(\theta)$ with the background, states as (\ref{entangled-boundary})
        actually describe solitons of the geometry with thermodynamic properties like black
         branes.

 \vspace{0.5cm}

 {\large\bf Temperature and dissipative effects}

 \vspace{0.5cm}

 It is important to stress that this entropy may, or may not, be associated to equilibrium
  states and temperature.
 However the construction above handles a model to describe the thermal properties of
  $D_p$-branes and closed string states being created on them.
 Let us consider the following potential,
\be
F = <H^a> - \frac1\beta \, S \sim (\tilde{F}) \,\,\, \,\, ,
\ee
and minimize it with respect to the transformation's parameters
$\theta$'s \cite{ume2}. Here $<H^a>$ is given by computing the
matrix elements of the right-handed Hamiltonian operator $H^a$ in the
state (\ref{tva}).
 The temperature is defined as the positive definite parameter $T\equiv (k_B\beta)^{-1}$
  considered independent of the variations of the parameters $\theta$.
The solution for the angular parameters $\theta_n$ is
given by the Bose-Einstein distribution: \be\label{bose} N_n (= \tilde{N}_n ) =\sinh^2
\theta_n = (e^{\b E_n } -1)^{-1 } .\ee
 The vacuum
state conditions (\ref{cond1}, \ref{cond2}), valued on these parameters $\theta_n (\b)$,
 provide the generalization of the boundary conditions at finite temperature. According
  to this formalism, a physical one-string excitation at temperature $T$, say $n,\tilde{n}$,
   may be created from this vacuum as usual:
\be \left.\left| 1; n ,\tilde{n},(\beta)\right\rangle \!\right\rangle\,\, = \,\,\frac{1}{n!}
\,\,a_{n}^{\dagger}(\b) \, \,{\tilde
a}_{n}^{\dagger}(\b)\,\,\left.\left|0_p(\beta)\right\rangle \!\right\rangle . \ee When the system
is in equilibrium, we see from (\ref{bose}) that entropy of the
bosonic brane goes to zero in the limit $T\rightarrow 0$. This
guarantees that the third principle of Thermodynamics is
satisfied. Notice, however, that there is an apparent paradox in
this model for thermal branes since the state (\ref{tva}) \emph{is
not} an eigenstate of the closed string Hamiltonian. Provided that
(\ref{exthamilt}) indeed describes the dynamics of the system, the
state (\ref{tva}) evolves and (\ref{bose}) cannot then be
considered an equilibrium distribution. Therefore, as a result, the
present model for thermodynamic effects is incomplete and
deviations from equilibrium should be included.

This may be seen as a notable coincidence with recent perspectives
on the hydrodynamic properties of thermal (black) branes
\cite{bb1,bb2,bb3}.
 According to these references, the infrared behavior of theories whose dual
bulk-gravities contain a black brane is governed by hydrodynamics,
and the main observation in this sense is the existence of a
universal value for the ratio of shear viscosity to entropy density
\cite{universal}, which should be investigated in the context of an
appropriate microscopical model.

Observe that the temperature in this model might be interpreted as due to
 the immersion of the brane in a thermal bath measured by accelerated
observers, according to the Unruh effect \cite{hor}. So, by identifying the
 temperature parameter with the relative acceleration of certain class of observers
  $a\equiv (2\pi k_B\beta)^{-1}$, this approach may also be viewed as describing
   accelerated branes \cite{russo}. In particular, for inertial observers, such that $a=0$,
    the standard string vacuum $|0\rangle$ in a Minkowski space-time is recovered, and the
     coherent state (\ref{entangled-boundary}) disappears, in agreement with the membrane paradigm \cite{paradigm}.

Let us finally mention that by virtue of the distribution (\ref{bose}),
 the divergence associated with the Hagedorn temperature is also present
  in this description, since the tension operator (\ref{tension}) tends to zero
   for higher temperatures, and thus the state
     $\,\,|0_p (\theta(\beta\to 0))\rangle \,\,$ is outside of the original Hilbert
      space in this limit \cite{nonunitarytfd}; i.e, the unitarity of the
       Bogoliubov transformation breaks down for sufficiently high temperatures
        \footnote{A consistent interpretation of this behavior was presented in Ref.
         \cite{hor} by studying closed strings crossing event horizons.}.

\section{Macroscopic pure states and collapse of closed string degrees of freedom}

The state (\ref{tva}) describes a statistical state or
 ensemble, however one could be interested in a sort of \emph{nonthermodynamic} limit
  in order to get a description of the D-brane as a macroscopic, but quantum-mechanical
   system\footnote{The simplest analogy is a \emph{perfect} crystal, an
   entropy-less macroscopic system}. This is a coherent state with vanishing entropy.

   In order to recover a quantum-mechanical picture, one usually studies
 the zero temperature limit, but there is a certain ambiguity in doing this.
  In particular by taking $\b\to\infty$ in (\ref{tva})
  one obtains the \emph{microscopic} vacuum,
   $\left|0\right\rangle$ of one-closed string.
    However more generally, if one takes the zero-entropy limit, the pure state that
      describes the system is recovered, and Nerst's theorem guarantees
     that $T \to 0$ as a particular way of taking such a limit (through a succession
      of equilibrium states).
         In this approach the entropy is an operator defined not only for equilibrium
          ensembles, and this limit may be realized preserving the macroscopic character
           expressed in the structure of coherent state. This is what we call the macroscopic
            pure state (MPS) limit.

As argued before, the state (\ref{tva}) may be writen in terms of the von Neumann entropy
 operator as follows
                 \be\label{boundarytermico-entropy2}
  \left.\left|0_p(\theta)\right\rangle\!\right\rangle=\, \frac{1}{C_{p}}\,e^{-K(\theta)/2}\, \,
\left|B_p \right\rangle \, ,\ee where
\be\label{pure} \left|B_p\right\rangle
\equiv{\cal B}_p\,\left.\left|0\right\rangle\!\right\rangle\equiv \, C_{p} \, \,\,\delta_p\,\prod_{I, n>0}
\,\, e^{ a^{\dagger\:I}_{n} \A^{\dagger\:I}_{n}} \left.\left|0\right\rangle\!\right\rangle \, .
\ee
This describes the macroscopic system as a branelike condensate, through the collective
 variables $\theta$, whose fundamental state corresponds to the fundamental equilibrium
  configuration given by the Bose-Einstein distribution (\ref{bose}).

 By projecting Eq. (\ref{boundarytermico-entropy}) in the
number basis $\left|n,\tilde{n}\right\rangle$ we have
\be\label{boundarytermico-proj} \left\langle n,\tilde{n}\left|0_{p}
(\theta) \right\rangle\!\right\rangle = \,e^{-\frac{S_n(\theta_n)}{2k_B}}\,
\langle n,\tilde{n} \left|B_p \right\rangle\, ,
\ee where $S_n(\theta_n)$ is the entropy of the $n^{th}$ level for
an arbitrary distribution (or collective state),
$\left\{\theta_n\right\}\equiv \left.\left|0_{p} (\theta)\right\rangle\!\right\rangle$.
Thus, we may observe the state $\left.\left|0_{p} (\theta)\right\rangle\!\right\rangle$
approaches $\left|B_p \right\rangle$ in the
zero-entropy limit $S_n(\theta_n)\to 0$\footnote{This limit may be
taken by choosing some arbitrary series of states
$\left\{\theta_n\right\}_M \equiv \theta_{n,\,M}$ with decreasing
entropies, $0 \leq S_n(\theta_{n,\,M})< S_n(\theta_{n,\,M-1}),
\forall \,\,n,\,M$.}, which may be identified with the coherent
pure state that describes the D-brane as a macroscopic
quantum-mechanical object, such as we aimed.

The microscopical (stringy) degrees of freedom described by the string
 canonical variables collapse in this limit. In fact, according to the expressions
  (\ref{tva}) and (\ref{boundarytermico-entropy}), this limit may be expressed by
   $\tanh \theta_n \to S_p (=\pm 1) $, or simply
\be\label{MPS} \left|(v_n/u_n)\right|^2 =\left|\tanh\theta_n\right|^2  \to 1 \, ;\ee
 thus if the canonical transformation is demanded to be unitary, the limit point cannot
  be reached unless the relation (\ref{Bmatrix}) is violated. By virtue of this, the
   canonical structure of the variables characterized by their CCR breaks down in this
    limit.
This is not surprising since the emerging macroscopic quantum-mechanical system must
 have proper (few) canonical degrees of freedom, instead of many stringy ones.
  Apart from this, as one also would expect for solitonic objects, despite the state of the system,
may even be represented in the string Fock space in this case (Eq. (\ref{boundarytermico-entropy})),
 its dynamics can no longer be described by perturbative degrees of freedom (strings).
     
Let us finally notice that the above macroscopic pure states may be also recovered
 from equilibrium states, when the temperature is analytically continued
  to purely imaginary values. This may be directly verified from the Bose-Einstein
   distribution: $ \tanh \theta_n = e^{n\b/2}$; by taking $\b\to i \tau$, the
    MPS condition $\left|\tanh\theta_n\right|^2= 1$ is fulfilled, or in other words,
     purely phase factors $ \tanh \theta_n $ may be absorbed in the creating operators
      of the state (\ref{tva}) through a canonical and unitary transformation. This
       remarkable property will be further explored elsewhere \cite{proximo}.

\vspace{0.5cm}

{\large\bf Canonical systems at finite temperature}

\vspace{0.5cm}

To end this work, we wish to point out briefly the implications of the ideas discussed
 above to very general thermodynamic systems since
the MPS limit may be a supplementary ingredient to study the
emergence of macroscopic states like branes in more diverse
contexts. In this sense the so-called thermofield dynamics (TFD) approach is a privileged
ground to do that.

Thermofield dynamics, developed by Takahashi
and Umezawa \cite{ume1,ume2,ume4,kha2,kha3},
is a real time approach to
quantum field theory at finite temperature \cite{kob1,leb1} where
an identical but fictitious copy of the system is introduced.
In TFD the full statistical
information of a quantum system is encoded in the (thermal) vacuum state instead
 of the density operator or partition function: \be\label{term-vac} 
  \left. \left| 0(\b )\right\rangle \!\right\rangle
 = Z^{-1 /2} \sum_n  e^{ -\b E_n /2 } | n \rangle |\tilde{n}
\rangle , \ee where $| n ; \tilde{n} \rangle$ denotes the $n^{th}$
energy eigenvalue of the two systems, the physical one and its
auxiliary copy denoted by $\,\,\tilde{}\,\,$. This may be
alternatively expressed in terms of a Bogoliubov transformation,
which map the Fock space based on the initial vacuum $| 0\rangle
\otimes |\tilde{0} \rangle$ (annihilated by $a^{}_{n}$ and
$\A^{}_{n}$) to a new thermal vacuum state: \be\label{vacTFD} 
\left. \left| 0(\theta )\right\rangle \!\right\rangle = e^{-i{G(\theta)}}  | 0 \rangle
|\tilde{0}\rangle . \ee The equilibrium state (\ref{term-vac})
corresponds to the particular distribution $ \tanh \theta_n =
e^{n\b/2}$, where the expectation value of the free energy
operator: ${\cal A}=H-\beta^{-1 } K  \,\,\,$ in the state
(\ref{vacTFD}), is stationary. The canonical entropy operator is
given here by \be K = -\,\sum_{n=1} \left[ a_{n}^{\dagger} a_{n }
\,\ln ( \sinh^{2}\theta _{n}) - a_{n} a_{n}^{\dagger} \, \ln (
\cosh^{2}\theta _{n})\right],\ee where $a^{\dagger}_{n}\,,\,\,
\A^{\dagger}_{n}$ refer to the canonical creation operators
corresponding to the physical system and its fictitious copy
respectively.

Consequently the concept of MPS limit may also be introduced in TFD,
 since one may write
\be\label{I-TFD}
  \left.\left|0(\theta)\right\rangle\!\right\rangle=\, \,e^{-K(\theta)/2}\, \,\prod_{ n>0}
\,\, e^{ a^{\dagger}_{n} \A^{\dagger}_{n}} \left.\left|0\right\rangle\!\right\rangle \,\,.\ee
 Thus the macroscopic pure states are
\be\label{pureTFD} \left.\left|I \right\rangle\!\right\rangle
\equiv{\cal I}\,\left.\left|0\right\rangle\!\right\rangle\equiv \, \, \,\prod_{ n>0}
\,\, e^{ a^{\dagger}_{n} \A^{\dagger}_{n}} \left.\left|0\right\rangle\!\right\rangle \, ,
\ee
which may also be obtained from the equilibrium ones (\ref{term-vac}) through analytic
 continuation of the temperature values.

\vspace{0.4cm}

\section{Final Remarks}

We built a statistical approach to a bosonic Dp-branes which handles a model to describe
 their thermodynamic properties, such that in the vanishing entropy limit the boundary
  states are recovered.  The picture is an ensemble of closed strings, which in the
   thermodynamic limit may be
seen as a sort of medium extended on $p$ spacial dimensions filled with string excitations.
 The model includes the evidence of dissipative behavior and the need for a hydrodynamic
  description \cite{bb1,bb2,bb3}. This radically differs from previous approaches to thermal
   D-branes using the TFD formalism \cite{IVV}, which supposes a duplication of the
    closed string degrees of freedom.
In future works we will investigate that (in)stability of the
thermal D-brane states and furthermore the possibility of studying black
branes using these ideas.

A remarkable strength of this description is that the $Dp$-branes are constructed
as the fundamental states of a closed string Fock space, which is consistent with the
 view of these objects as solitonic gravitational backgrounds.
 
 This work may be considered as a previous step addressed to formulate an open/closed dictionary,
  which should be based on the consistency with the description of thermal branes in terms of open strings \cite{{brane-tfd}}.

\section{Acknowledgements}
The author would like to thank to N. Grandi, J. A. Helayel-Neto and R. Scherer Santos
 for useful comments and observations on
the subject of this paper. This work
was partially supported by ANPCyT (Grant No. PICT-
2007-00849).

\end{document}